\documentclass[pra,aps,10pt,twocolumn,superscriptaddress]{revtex4} 
\usepackage{amsmath}
\usepackage{amssymb}
\usepackage{amsthm}
\usepackage{mathrsfs}
\usepackage{amsfonts}
\usepackage{enumerate}
\usepackage{latexsym}
\usepackage{bm}
\usepackage{graphicx}
\usepackage{subfigure}
\usepackage{booktabs}
\usepackage[dvipsnames]{xcolor}
\usepackage[colorlinks]{hyperref}
\hypersetup{
 colorlinks,
 citecolor=Violet,
 linkcolor=Red,
 urlcolor=Blue}
\usepackage{multirow}
\usepackage[normalem]{ulem} 
\definecolor{awesome}{rgb}{1.0, 0.13, 0.32}
\definecolor{electricpurple}{rgb}{0.75, 0.0, 1.0}
\definecolor{crimson}{rgb}{0.86, 0.08, 0.24}

\newcommand{\beq}{\begin{equation}}
\newcommand{\eneq}{\end{equation}}
\input{epsf}

\begin{document}

\title{Spectral Signatures of the Markovian to Non-Markovian Transition in Open Quantum Systems
}

\author{Zeng-Zhao Li}
\email{zengzhaoli09@gmail.com}
\affiliation{School of Science, Harbin Institute of Technology, Shenzhen, China}

\author{Chi-Hang Lam}
\affiliation{Department of Applied Physics, Hong Kong Polytechnic University, Hung Hom, Hong Kong, China}

\author{Cho-Tung Yip}
\email{h0260416@hit.edu.cn}
\affiliation{School of Science, Harbin Institute of Technology, Shenzhen, China}

\author{Bo Li}
\email{libo2021@hit.edu.cn}
\affiliation{School of Science, Harbin Institute of Technology, Shenzhen, China}

\begin{abstract}
We present a new approach for investigating the Markovian to non-Markovian transition in quantum aggregates strongly coupled to a vibrational bath through the analysis of linear absorption spectra. Utilizing hierarchical algebraic equations in the frequency domain, we elucidate how these spectra can effectively reveal transitions between Markovian and non-Markovian regimes, driven by the complex interplay of dissipation, aggregate-bath coupling, and intra-aggregate dipole-dipole interactions. Our results demonstrate that reduced dissipation induces spectral peak splitting, signaling the emergence of bath-induced non-Markovian effects. 
The spectral peak splitting can also be driven by enhanced dipole-dipole interactions, although the underlying mechanism differs from that of dissipation-induced splitting.  
Additionally, with an increase in aggregate-bath coupling strength, initially symmetric or asymmetric peaks with varying spectral amplitudes may merge under weak dipole-dipole interactions, whereas strong dipole-dipole interactions are more likely to cause peak splitting. 
Moreover, we find that spectral features serve as highly sensitive indicators for distinguishing the geometric structures of aggregates, while also unveiling the critical role geometry plays in shaping non-Markovian behavior.  
This study not only deepens our understanding of the Markovian to non-Markovian transition but also provides a robust framework for optimizing and controlling quantum systems. 
\end{abstract}

\date{\today}
\maketitle

\section{Introduction}

The study of non-Markovian effects in open quantum systems has garnered significant attention in quantum optics, quantum information, and quantum chemistry~\cite{ScullyZubairy97,NielsenChuang00,Breuer16rmp,Vega17rmp,May11}. These effects provide deep insights into the fundamental interactions between quantum systems and their environments, shedding light on complex coherence and decoherence dynamics essential for advancing quantum technologies like quantum computing, communication, and sensing. However, experimental detection and precise control of non-Markovian processes remain challenging due to the complexity of environmental interactions. Overcoming these challenges can enable innovative strategies for error correction and optimization in quantum devices, enhancing their reliability and performance~\cite{BallBiercuk16pra,WhiteHollenbergModi20ncommun,Hakoshima21pra,Cai23rmp,Ahn24aqt,RossiniDonvil23prl}.

Recent experiments have demonstrated Markovian to non-Markovian transitions in quantum optical systems~\cite{LiuLiBreuerPiilo11nphys} and explored non-Markovian effects in diverse contexts, including entanglement-assisted probing~\cite{GaikwadMurch24prl}, entanglement oscillations~\cite{CialdiParis11pra}, measurements of non-Markovianity~\cite{TangLiBreuerPiilo12epl,Bernardes15srep}, micromechanical Brownian motion~\cite{Groblacher15ncomms}, and multi-exponential decay dynamics in quantum dots~\cite{Madsen11prl}. These advances underscore the need for theoretical frameworks capable of describing non-Markovian dynamics efficiently. Key approaches include master equations~\cite{Breuer02}, non-Markovian stochastic Schrödinger equations~\cite{GardinerZoller00,DiosiStrunz1997PLA,DiosiGisinStrunz98pra}, path integral methods~\cite{Feynman48rmp,Feynman63annphys,CaldeiraLeggett83annphys,CaldeiraLeggett1983Physica,Weiss08,Makri95jmp,Makri18JCP,Zhang12prl}, quantum jump methods~\cite{Piilo08prl}, hierarchical equations of motion~\cite{TanimuraKubo1989jpsj,Tanimura1990pra,Tanimura20JCP}, and tensor network techniques~\cite{VerstraeteCirac04prl,Daley14AdvPhys,StrathearnKeelingLovett18ncommun}. 
Among stochastic methods, non-Markovian quantum state diffusion (NMQSD) equation, a specialized stochastic Schrödinger equation, is a versatile tool for studying diverse open quantum systems. Unlike exact master equations, which are limited to specific models~\cite{Hu92prd,Strunz99prl,Yu99pra}, NMQSD provides an exact and broadly applicable formalism~\cite{Yu04pra, JingYu10prl, LiYouLam14pra, LuoYou15pra}. 
The hierarchy of pure states (HOPS) approach~\cite{SussEisfeldStrunz14prl} extends the NMQSD framework by resolving functional derivatives, offering a systematic representation of non-Markovian effects via a hierarchy of pure states. This method has been further refined in recent studies~\cite{HartmannStrunz17JCTC, HartmannStrunz21JPCA, Gao22PRA, GeraRaccah23JCP, CittyRaccah24JCP}. Building on NMQSD and HOPS, hierarchical equations in the time~\cite{RitschelEisfeld15jcp} and frequency domains~\cite{ZhangLiEisfeld17} have been developed to describe absorption spectra of molecular aggregates strongly coupled to vibrational modes~\cite{WitkowskiMoffitt60, FultonGouterman60, SchererFischer84, Eisfeld06}.

Quantum aggregates, such as molecular systems, exhibit collective quantum phenomena, including coherence and excitonic states, distinct from their individual components~\cite{Davydov13}. These aggregates demonstrate behaviors like superradiance and tunneling, crucial for applications in quantum computing~\cite{Scholes10jpcl}, photosynthesis~\cite{Scholes2011ncehm}, and nanotechnology~\cite{GorlWurthner12}. They also provide a versatile platform for exploring fundamental aspects of quantum mechanics~\cite{HestandSpano18chemrev}. 
By analyzing their absorption spectra, one can uncover essential features like memory effects, non-Markovian dynamics, and energy transfer processes, providing key insights into their electronic structure, coherence, and interactions with the environment. 

Previous studies have examined non-Markovian effects in absorption spectra across various contexts. Roden {\it et al.} combined the NMQSD framework with an abstract approximation to calculate energy transfer and absorption spectra of molecular aggregates, comparing the results with exact pseudomode methods to validate their approach~\cite{RodenStrunzEisfeld11JCP}. Ritschel {\it et al.} utilized NMQSD alongside the thermofield method to compute temperature-dependent optical spectra of light-harvesting aggregates, effectively incorporating environmental vibrations~\cite{RitschelEisfeld15jcp}. Schröder {\it et al.} demonstrated how bath spectral densities and long correlation times shape absorption line shapes~\cite{Schröder06JCP}, while Vogel {\it et al.} employed femtosecond absorption spectroscopy to reveal non-Markovian correlations and hole-burning phenomena~\cite{Vogel88PRA}. 

Building on these advancements, our study establishes the absorption spectrum as a direct and experimentally accessible tool for investigating the Markovian to non-Markovian transition in open quantum systems. Unlike prior studies focusing on specific systems or temporal correlations, we systematically connect spectral features—such as peak splitting, merging, and shifting—to dissipation, aggregate-bath coupling, and geometry, providing a unified framework for understanding non-Markovian behavior. Using hierarchical equations in the frequency domain~\cite{ZhangLiEisfeld17}, we demonstrate how reduced dissipation induces peak splitting, signaling a transition to non-Markovianity, while enhanced dipole-dipole interactions and aggregate-bath coupling create distinct spectral patterns. By analyzing configurations from monomers to tetramers, we reveal how geometry influences non-Markovian effects, with linear aggregates exhibiting pronounced spectral variations compared to the stability of symmetric ring geometries. These findings establish the absorption spectrum as a powerful tool for studying non-Markovian effects and advancing our understanding of quantum coherence and energy transfer processes.
While conventional measures such as the trace distance criterion are widely used to characterize non-Markovianity, they often abstract the underlying physical mechanisms. In contrast, our spectral approach provides a complementary perspective by linking memory effects to observable spectral features. As discussed in detail later, this connection offers both physical insight and experimental accessibility.

The paper is organized as follows. Sec.~\ref{sec:model} introduces the model for a quantum aggregate. In Sec.~\ref{sec:method}, we describe the hierarchical equations in the frequency domain used to compute the absorption spectrum.  Sec.~\ref{sec:result} presents the absorption spectra, revealing transitions between Markovian and non-Markovian regimes, as well as the distinctive geometric features of the aggregates. Finally, Sec.~\ref{sec:conclusion} offers a discussion of the findings and concludes the paper.

\begin{figure}
\centering
\includegraphics[width=1\columnwidth]{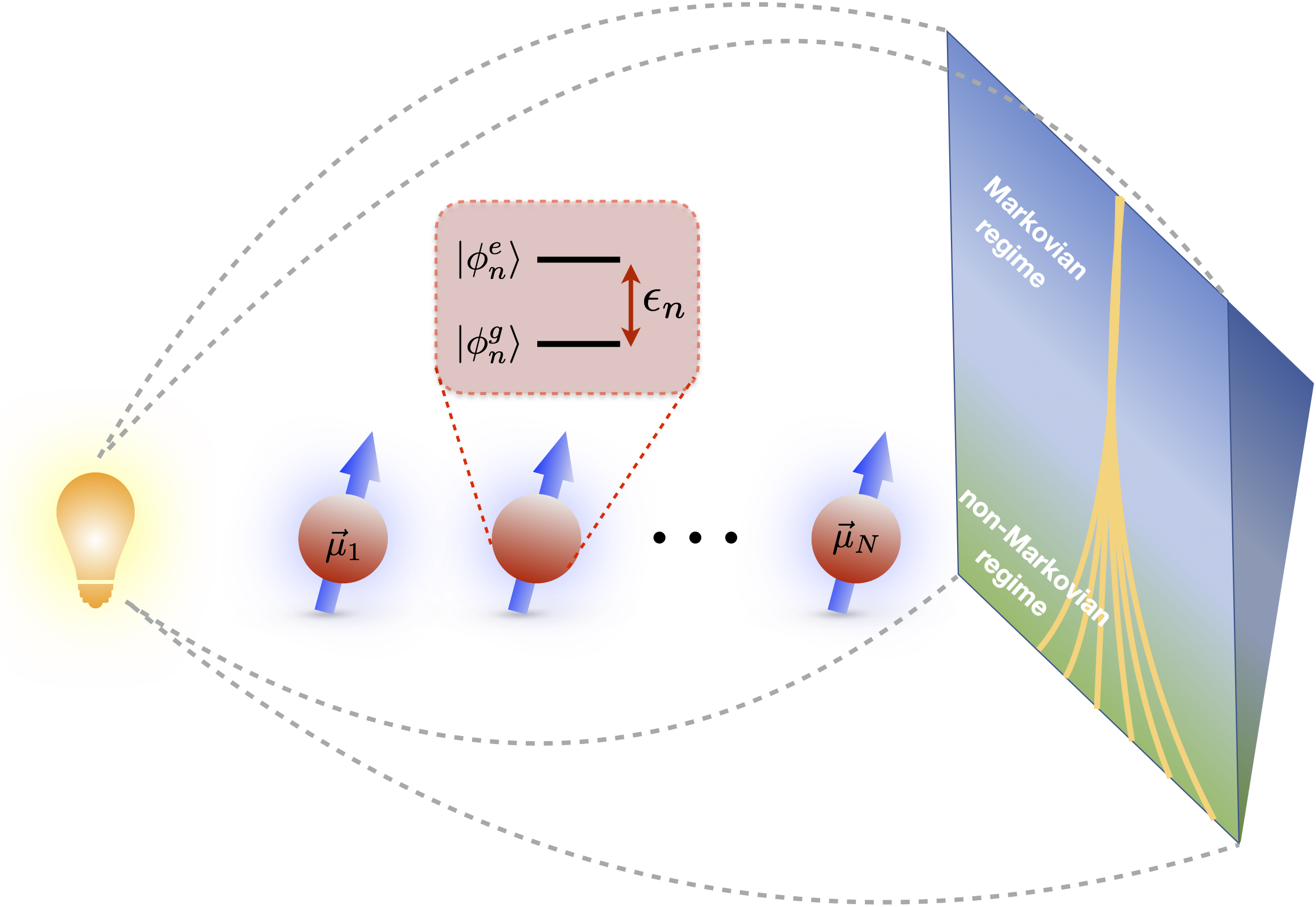} 
\caption{Illustration of a quantum aggregate system with $N$ two-level monomers, each represented by a sphere with a dipole moment indicated by an arrow. The surrounding blue cloud represents the vibrational bath interacting with each monomer. The insert shows the electronic states for a single monomer, where $|\phi_n^g\rangle$ and $|\phi_n^e\rangle$ denote the ground and excited states, respectively, with an energy separation $\epsilon_n$. The dipoles are coupled through dipole-dipole interactions, influencing the excitation dynamics across the aggregate. A light source on the left excites the system, while the spectrum detector on the right captures the transition dynamics, illustrating the shift from Markovian to non-Markovian regimes.
}
\label{fig:schematic}
\end{figure}

\section{Quantum Aggregate Model \label{sec:model}}

We consider a quantum aggregate consisting of $N$ two-level monomers, each of which is strongly coupled to its own vibrational bath, as illustrated in Fig.~\ref{fig:schematic}. In the single-excitation manifold, where only one monomer is allowed to be excited at any given time, the many-body ground and excited states are defined as $
\lvert g_{\rm el}\rangle=\prod_{m=1}^{N} \lvert\phi_m^{g}\rangle$ and $
\lvert \pi_n\rangle=\lvert\phi_n^{e}\rangle\prod_{m\neq n}^N \lvert\phi_m^g\rangle , 
$
respectively. Here $\lvert\phi_n^{g}\rangle$ denotes the ground state of the $n$th monomer, and $\lvert\phi_n^{e}\rangle$ represents the excited state. 
 
The total Hamiltonian in the single-excitation subspace is expressed as \cite{RitschelEisfeld15jcp} 
\begin{eqnarray}
H&=&H_{\rm sys}+H_{\rm env}+H_{\rm int},
\end{eqnarray}
where
\begin{eqnarray}
H_{\rm sys}&=&\sum_{n=1}^{N} \epsilon_n \lvert\pi_n\rangle\langle \pi_n\rvert +\sum_{\substack{n,m=1\\n\neq m}}^{N} V_{nm} \lvert\pi_n\rangle\langle\pi_m\rvert, \label{eq:sys} \\
H_{\rm env}&=&\sum_{n=1}^N\sum_{\lambda} \omega_{n\lambda}a^{\dagger}_{n\lambda} a_{n\lambda}, \\
H_{\rm int}&=&-\sum_{n=1}^N L_n \sum_{\lambda} \kappa_{n\lambda} (a_{n\lambda}^{\dagger}+a_{n\lambda}) .
\end{eqnarray}
In the system Hamiltonian $H_{\rm sys}$, $\epsilon_n$ denotes the excitation energy of the $n$th monomer ($n=1,\cdots,N$) and $V_{nm}$ represents the dipole-dipole interaction between the monomers $n$ and $m$. For this work, we consider degenerate levels in each monomer, setting $\epsilon_n=0$~\cite{RitschelEisfeld15jcp}.  
If $\lvert n-m\rvert=1$, the dipole-dipole interaction is given by $V_{nm}=V\cos\theta$, where $\theta$ is the angle between the dipole moments $\vec{\mu}_n$ and $\vec{\mu}_m$~\cite{LiEisfeld17pra}. For all other cases, $V_{nm}=0$. 
In the interaction Hamiltonian $H_{\rm int}$, $L_n=\lvert\pi_n\rangle\langle\pi_n\rvert$ denotes the system operator linearly coupled to the bath modes.

In the environmental Hamiltonian $H_{\rm env}$, $a_{n\lambda}^{\dagger}$ and $a_{n\lambda}$ represent bosonic operators for the vibrational modes of the bath. The coupling strength between the aggregate and the vibrational modes is characterized by $\kappa_{n\lambda}$ in the interaction Hamiltonian $H_{\rm int}$. Both $\kappa_{n\lambda}$ and $a_{n\lambda}^{\dagger}$ are indexed by $n$, indicating that each monomer in the quantum aggregate interacts with an independent environment. This coupling strength is typically encoded in a spectral density $J_n(\omega)=\pi\sum_k \lvert \kappa_{n\lambda}\rvert^2\delta(\omega-\omega_{n\lambda}).$ 

For simplicity, we assume Orstein-Uhlenbeck noise~\cite{UhlenbeckOrnstein1930,Gardiner}, which is utilized in the development of the NMQSD~\cite{Strunz99prl,Yu99pra,JingYu10prl,LiYouLam14pra,LuoYou15pra}. The corresponding bath correlation function %
and the Lorentzian bath spectrum at zero temperature are described as 
\begin{eqnarray}
\alpha_n(\tau)&=&g e^{iw_{n}\tau} ~~~~~ (\text{for }\tau>0) ,  
\label{eq:bcf}
\end{eqnarray}
and
\begin{eqnarray}
J_n(\omega)&=&\int_{-\infty}^{\infty} d\tau  \alpha_n(\tau) e^{-i\omega \tau} =\frac{2g\gamma}{\gamma^2+(\omega-\Omega)^2} .
\label{eq:spectral_density}
\end{eqnarray}
Here $w_{n}=\Omega_n+i\gamma_n$ with $\Omega_n=\Omega$ and $\gamma_n=\gamma$ implying each monomer experiences a similar but independent environment. 
The spectral density $J(\omega)$ in Eq.~(\ref{eq:spectral_density}) describes how the environment's different frequency modes $\omega$ contribute to the system's dynamics. The parameter $g$ represents the coupling strength between the aggregate and the bath, while the correlation parameter $\gamma$ accounts for the bath-induced dissipation rate of the quantum aggregate. For consistency in our calculations, we use $\Omega = 1$ as the energy unit.

\section{Laplace-domain hierarchy of equations \label{sec:method}} 

The conventional method for calculating the linear absorption spectrum involves solving the system dynamics in the time domain and then applying a Fourier transform to the dipole-dipole correlation function~\cite{May11,RitschelEisfeld15jcp}. In contrast, our work adopts a Laplace-domain approach, deriving a hierarchy of linear algebraic equations directly in the Laplace domain. This differs from the traditional time-domain hierarchical equations, which are based on differential equations~\cite{ZhangLiEisfeld17}.

The linear absorption spectrum is defined as the half-sided Fourier transform of the dipole-dipole correlation function~\cite{RitschelEisfeld15jcp,May11}, expressed as $F(\omega) =\sum_{n,m} \vec{\mu}_n \vec{\mu}_m \Re[\int_0^{\infty} dt e^{i\omega t} C_{nm}(t)]$, where $\vec{\mu}_n$ represents the transition dipole moment of the $n$th monomer and $C_{nm}(t)=\langle \pi_n|\psi_m^{(\vec{0})}(t) \rangle$. The state $\vert\psi_m^{(\vec{0})}(t) \rangle$ is obtained by solving a hierarchy of coupled states $\vert\psi^{(\vec{k})}(t)\rangle$ as governed by Eq.~(\ref{eq:hops_linear}) with the initial condition $\vert\psi^{(\vec{0})}(t=0)\rangle=|\pi_m\rangle$, indicating that the excitation is initially localized in the $m$th monomer. 

This absorption spectrum can be reformulated using the Laplace transformation as 
\begin{eqnarray}
F(\omega) 
&=& \sum_{n,m} \vec{\mu}_n \vec{\mu}_m \Re[  \lim_{\epsilon\rightarrow0} \tilde{C}_{nm}(s) ], 
\label{eq:spectrum}
\end{eqnarray}
where $\vec{\mu}_n \vec{\mu}_m=\mu^2$ for parallel dipoles and 
$\tilde{C}_{nm}(s)=\mathscr{L}[C_{nm}(t)]
=\int_0^{\infty}dt e^{-st} C_{nm}(t)$ 
with $s=\epsilon-i\omega$, where both $\epsilon$ and $\omega$ are real. From the definition of $\mathcal{C}_{nm}(t)$, it follows that 
\begin{eqnarray}
	\mathcal{\tilde{C}}_{nm}(s) &=& \langle \pi_n|\Psi_m^{\vec{(0)}}(s) \rangle , 
\label{eq:tilde_C}
\end{eqnarray}
and the hierarchy of coupled algebraic equations for $|\Psi^{\vec{(k)}}(s) \rangle$ %
is given by (see Appendix~\ref{sec:AppendA}) 
\begin{eqnarray}
\vert\psi^{(\vec{k})}(t=0)\rangle 
&=&
(s+iH_{\rm sys} + i\sum_{n}k_{n}\omega_{n}) \vert\Psi^{(\vec{k})}(s)\rangle \notag\\
&&-\sum_{n}L_n k_{n}\alpha_{n}(0) \vert\Psi^{(\vec{k}-\vec{e}_{n})}(s)\rangle \notag\\
&&+\sum_{n}L_n^{\dagger} \vert\Psi^{(\vec{k}+\vec{e}_{n})}(s)\rangle . 
\label{eq:hops_linear_laplace_nonoise}
\end{eqnarray}
Here, $\vec{k}=(k_{1},\cdots,k_{N})$ 
with $k_{n}\ge 0$ being integers, and $\vec{e}_{n}=\{0,\cdots,1,\cdots,0\}$ is a vector with a single nonzero value of $1$ at the $n$th position. The truncation scheme considers only $\vec{k}$ values satisfying $\sum_{n}k_{n}\le E_{\rm max}$, where $E_{\rm max}$ is chosen to be sufficiently large to ensure convergence. 
To guarantee the convergence of the hierarchical expansion, we set the truncation level to $E_{\rm max} = 12$, which has been verified to be adequate across all parameter regimes considered in this study. 
Alternative truncation schemes are discussed in Refs.~\onlinecite{SussEisfeldStrunz14prl} and \onlinecite{ZhangEisfeld18jcp}. In the following Section~\ref{sec:result}, we utilize Eqs.~(\ref{eq:spectrum}) and (\ref{eq:tilde_C}) along with the solution of Eq.~(\ref{eq:hops_linear_laplace_nonoise}) to calculate the linear absorption spectrum of a quantum aggregate.

The numerical simulations in this work are based on the zeroth-order stochastic equation [as given in Eq.~(\ref{eq:hops_linear_laplace_nonoise})], which provides an efficient approximation to capture the influence of system-bath interactions. While this approach is formally most accurate under weak to intermediate coupling and finite bath correlation times, it has been widely demonstrated to yield qualitatively reliable results even in regimes where non-Markovian effects are significant~\cite{TanimuraKubo1989jpsj,IshizakiTanimura05JPSJ}. In particular, for the calculation of linear absorption spectra, the zeroth-order approximation effectively captures the essential spectral features arising from dissipation and bath-induced correlations~\cite{RitschelEisfeld15jcp}. In our simulations, we have verified the numerical stability and convergence of results across a broad parameter range, including small $\gamma$ and moderately strong $g$, and have not observed any unphysical behavior, supporting the applicability of this approximation in the present study.

We note that since the calculations are based on hierarchical algebraic equations in the frequency domain, time is not explicitly considered in the simulations. The frequency-domain formulation is obtained via Laplace transformation of the time-domain equations, which, in principle, involves integration over an infinite time interval. This ensures that the full steady-state response of the system is captured, provided that the relevant timescales of system dynamics are well encompassed. Therefore, explicit time truncation is not required in this framework.

\section{Non-Markovian absorption spectrum \label{sec:result}}

In this section, we analyze the absorption spectra of various quantum aggregates, including monomers, dimers, and more complex trimer and tetramer configurations in both linear and ring geometries. We demonstrate that monomers and dimers exhibit fundamental signatures of the Markovian to non-Markovian transition. Furthermore, our analysis of trimers and tetramers reveals that specific spectral features can accurately identify and differentiate the geometric structures of these aggregates, providing deeper insights into the manifestation of non-Markovian effects.

\subsection{Analytical Solutions for Specific Limiting Cases \label{sec:anaylitics}}

We first consider some analytically solvable limiting cases. When the aggregate is decoupled from the vibrational environment ($g = 0$) or when environmental dissipation is overwhelmingly strong (e.g., $\epsilon_n,\,V_{nm} \ll \gamma \rightarrow \infty$), the spectral properties can be accurately derived by focusing solely on the system Hamiltonian $H_{\rm sys}$, which allows for exact diagonalization. In these cases, spectral amplitudes are generally more pronounced in the absence of environmental coupling compared to when dissipation is dominant. Here, the system dynamics are predominantly determined by the eigenenergies and eigenstates of the Hamiltonian $H_{\rm sys}$, either due to the absence of coupling to the vibrational bath or because the bath dynamics are sufficiently fast to reach a steady state before the system evolves (e.g., $\gamma \gg \epsilon_n,\,V_{nm}$). Consequently, these eigenenergies manifest as spectral peaks, which correspond to the electronic transitions within the aggregate. 

For a {\it linear} aggregate with open boundary conditions, the eigenfunctions of the system Hamiltonian in Eq.~(\ref{eq:sys}) are described by~\cite{Malyshev95prb} 
\begin{eqnarray}
\lvert \psi_j^{\rm (L)}\rangle=\sum_{n=1}^{N} c_{jn}^{\rm (L)}\lvert n\rangle 
\end{eqnarray}
with $N\geq j \geq 1$ and 
\begin{eqnarray}
c_{jn}^{\rm (L)}=  \sqrt{\frac{2}{N+1}} \sin(\frac{jn\pi}{N+1}) .
\end{eqnarray}
The corresponding eigenvalues $\omega_j^{\rm (L)}$ and oscillator strengths $f_j^{\rm (L)}$ (which are proportional to the spectral amplitudes~\cite{AtkinsFriedman10,Green20}) for absorption from the electronic ground state to an excited state are given by 
\begin{eqnarray}
\omega_j^{\rm (L)}&=&2V\cos\theta\cos\Big(\frac{j\pi}{N+1}\Big),  \label{eq:1D_eigenvalue} \\ 
f_j^{\rm (L)}&=& \Big\lvert \sum_{n=1}^{N}c^{\rm (L)}_{jn}\Big\rvert^2=\frac{1-(-1)^j}{N+1} \cot^2\Big(\frac{j\pi}{2(N+1)}\Big). \label{eq:1D_oscillatorStrength}
\end{eqnarray}
Spectral peaks with strengths $f_j^{\rm (L)}$ (where the oscillator strength of a monomer is normalized to unity) are expected at energies $\omega_j^{\rm (L)}$. Using Eq.~(\ref{eq:1D_eigenvalue}), the transition frequencies $\omega_j^{\rm (L)}$ are calculated and listed in Table~\ref{table:obc}. According to Eq.~(\ref{eq:1D_oscillatorStrength}), only transitions with odd values of $j$ are observable in the spectrum and these transitions are represented by frequencies displayed in Table~\ref{table:obc}.

\begin{table}[ht]
\caption{Electronic transition frequencies of linear quantum aggregates. Only transitions with frequencies $\omega_j^{(L)}$ with odd $j$ ($N\geq j\geq 1$) are observable in the absorption spectrum. The parameters used are $V=1$ in units of $\Omega$ and $\theta=0$.} 
\centering 
\begin{tabular}{c c c c c} 
\hline\hline 
$N$ & $\omega_{j=1}^{\rm (L)}$ & $\omega_{j=2}^{\rm (L)}$ & $\omega_{j=3}^{\rm (L)}$ & $\omega_{j=4}^{\rm (L)}$ \\ [0.5ex] 
\hline
$1$ & $0$ &   & & \\ 
$2$ & $1$ &   $-1$ &  & \\
$3$ & $\sqrt{2}$ &  $0$ & $-\sqrt{2}$ & \\
$4$ & $\frac{\sqrt{5}+1}{2}$ & $\frac{\sqrt{5}-1}{2}$ & $-\frac{\sqrt{5}-1}{2}$ &  $-\frac{\sqrt{5}+1}{2}$  \\ [1ex] 
\hline 
\end{tabular}
\label{table:obc}
\end{table}

For a {\it ring} aggregate with periodic boundary conditions, the eigenfunctions of the system Hamiltonian in Eq.~(\ref{eq:sys}) with $|\pi_{N+n}\rangle = |\pi_n\rangle$ are expressed as~\cite{Malyshev95prb} 
\begin{eqnarray}
\lvert \psi_j^{\rm (C)}\rangle=\sum_{n=1}^{N} c_{jn}^{\rm (C)}\lvert n\rangle , 
\end{eqnarray}
where  
\begin{eqnarray}
c_{jn}^{\rm (C)}= \frac{1}{\sqrt{N}} e^{i\frac{2jn\pi}{N}} .
\end{eqnarray}
The corresponding transition frequencies and oscillator strengths are given by
\begin{eqnarray}
\omega_j^{\rm (C)}&=&2V\cos\theta\cos\Big(\frac{2j\pi}{N}\Big),  \label{eq:2D_eigenvalue} \\
f_j^{\rm (C)}&=& \Big\lvert \sum_{n=1}^{N}c_{jn}^{\rm (C)}\Big\rvert^2 =\frac{(N-1)^2}{N} \delta_{jN}.  \label{eq:2D_oscillatorStrength}
\end{eqnarray}
Eqs.~(\ref{eq:2D_eigenvalue}) and (\ref{eq:2D_oscillatorStrength}) indicate that only a single observable peak at $\omega_N^{\rm (C)}$ is present in the spectrum, as shown as $j=N$ in Table~\ref{table:pbc}.

\begin{table}[ht]
\caption{Electronic transition frequencies of ring aggregates. Only transitions with frequencies $\omega_j^{(C)}$ with $j=N$ can be observed in the absorption spectrum. The parameters used are $V=1$ in units of $\Omega$ and $\theta=0$.} 
\centering 
\begin{tabular}{c c c c c} 
\hline\hline 
$N$ & $\omega_{j=1}^{\rm (C)}$ & $\omega_{j=2}^{\rm (C)}$ & $\omega_{j=3}^{\rm (C)}$ & $\omega_{j=4}^{\rm (C)}$ \\ [0.5ex] 
\hline 
$3$ & $-1$ & $-1$ & $2$ & \\
$4$ & $0$ & $-2$ & $0$ & $2$ \\ [1ex] 
\hline 
\end{tabular}
\label{table:pbc}  
\end{table}

In particular, when the dipole-dipole interaction is absent (i.e., $V = 0$), the spectral peak position for an aggregate decoupled from the vibrational bath ($g=0$) coincides with $\omega_j^{\rm (L)} = \omega_j^{\rm (C)} = 0$, as given by Eqs.~(\ref{eq:1D_eigenvalue}) and (\ref{eq:2D_eigenvalue}), identical to that of a single monomer.

Note that in this subsection, we explore the oscillator strength, which is directly proportional to the spectral amplitude of the absorption spectrum~\cite{AtkinsFriedman10,Green20}.  
In the subsequent Subsections~\ref{IVB} and \ref{IVC}, we will focus on analyzing the spectral amplitude obtained from numerical results.

\subsection{Spectral Signatures of the Markovian to Non-Markovian Transition}
\label{IVB}

\subsubsection{Spectral analysis of a monomer}

\begin{figure}
\centering
  \includegraphics[width=1\columnwidth]{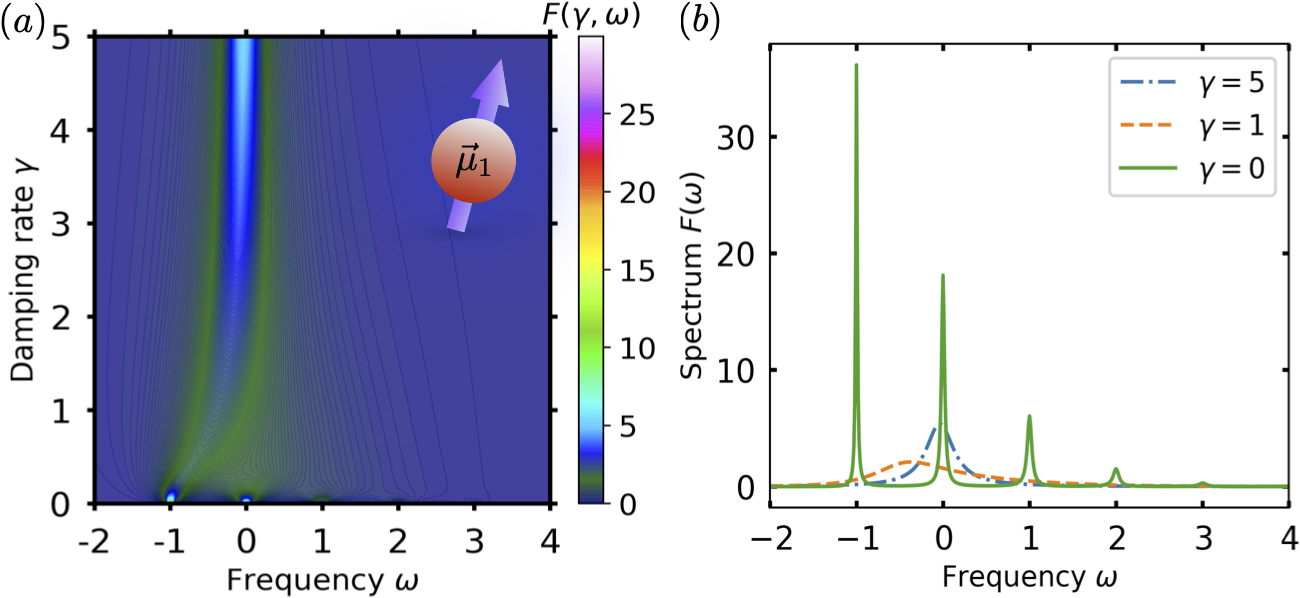} 
\caption{(color online) (a) Absorption spectrum $F(\gamma,\omega)$ of a monomer as a function of the dissipation rate $\gamma$. (b) One-dimensional spectra extracted from (a) for specific dissipation rates: $\gamma=0,\,1,\,5$. With the monomer's electronic transition frequency set to $\omega_0=0$, a spectral peak emerges at large $\gamma$ with $\omega \approx \omega_1^{\rm (L)} = 0$, consistent with Eq. (\ref{eq:1D_eigenvalue}). For $\gamma = 0$, these peaks are located at $\omega = \omega_1^{\rm (L)} = -1,\,0,1,\,2,\,3,\cdots$. Spectra are presented in units of $\mu^2$. The hierarchy depth is set to 12, with additional parameters $g = 1$ and $\Omega = 1$.}
\label{fig:fig_monomer}
\end{figure}

Figure~\ref{fig:fig_monomer}(a) displays the linear absorption spectrum $F(\gamma,\omega)$ of a monomer [illustrated in the inset of Fig.~\ref{fig:fig_monomer}(a)] coupled to the environment. As expected a single peak at $\omega \approx \omega_1^{\rm (L)}=0$ (see Table~\ref{table:obc} with $j=N=1$), representing an electronic transition of a monomer, is observed at strong dissipation corresponding to a large $\gamma$. 
As $\gamma$ decreases, the peak broadens and splits, becoming evident when the distance between the peaks surpasses the width of each individual peak. This occurs around $\gamma \approx 1$, indicating a bath-induced transition in the monomer. 
In particular, for $\gamma=0$, the spectral peaks are exactly located at $\omega=-1,\,0,\, 1,\, 2,\, 3,\, \cdots$ with decreasing amplitudes~\cite{May11,Roden11JCP}. 
These observations are further illustrated in the one-dimensional spectra in Fig.~\ref{fig:fig_monomer}(b) for three different values of the dissipation rate, namely, $\gamma =0,\,1,\,5$.  
The characteristic time of a monomer can be expressed as $1/\omega_0$, where $\omega_0=\omega_1^{\rm (L)}$. In the current scenario, $\omega_0=0$ due to the consideration of $\epsilon_n=0$ in Eq.~(\ref{eq:sys}).
Physically, if the bath correlation time  $1/\gamma$ [see Eqs.~(\ref{eq:bcf}) and (\ref{eq:spectral_density})] is larger than or comparable to the monomer characteristic time $1/\omega_0$, strong non-Markovian effects induced by the vibrational bath modes on the electronic states will be present. Then, bath-induced transitions become dominant, leading to the appearance of multiple peaks in the spectrum. 
In contrast, if the system dynamics are slower than those of the bath, such as in cases of strong dissipation ($\gamma \gg \omega_0$) where the bath reaches its steady state before the system evolves, the system will be in the Markovian regime. This results in a single peak at $\omega = \omega_1^{\rm (L)}=0$ for the electronic transition of a monomer. Consequently, the spectrum of the monomer distinctly reveals the transition from Markovian to non-Markovian behavior as environmental dissipation is reduced. 
An alternative but fundamentally equivalent interpretation can be explained in the following way. 
When the correlation parameter $\gamma$ is small (weak dissipation), information lost to the environment can flow back into the system, leading to memory effects and, consequently, a non-Markovian environment. In contrast, when $\gamma$ is large (strong dissipation), information rapidly dissipates into the environment and cannot return to the system, rendering the environment memoryless and Markovian.

We analyze the spectral density $J_n(\omega)$, defined as the Fourier transform of the bath correlation function $\alpha_n(\tau)$ in Eq.~(\ref{eq:spectral_density}). Setting $\gamma = 0$ reduces $J_n(\omega)$ to a delta function, $J_n(\omega) = 2 \pi g \delta(\omega - \Omega)$, indicating a spectrum sharply peaked at a single frequency $\Omega$. This describes a purely coherent, non-dissipative environment, where bath modes are confined to a single frequency with no broadening, resulting in no energy dissipation. Consequently, $\alpha_n(\tau)$ becomes purely oscillatory, indicating no decay and suggesting energy retention within the system-bath interaction. This idealized case isolates vibrational structure effects in a coherent setting, highlighting non-dissipative conditions. For a more realistic model, a nonzero $\gamma$ would introduce spectral broadening and richer non-Markovian effects by enabling interactions across a range of frequencies, thus yielding more complex dynamics.

\begin{figure}
\centering
  \includegraphics[width=1\columnwidth]{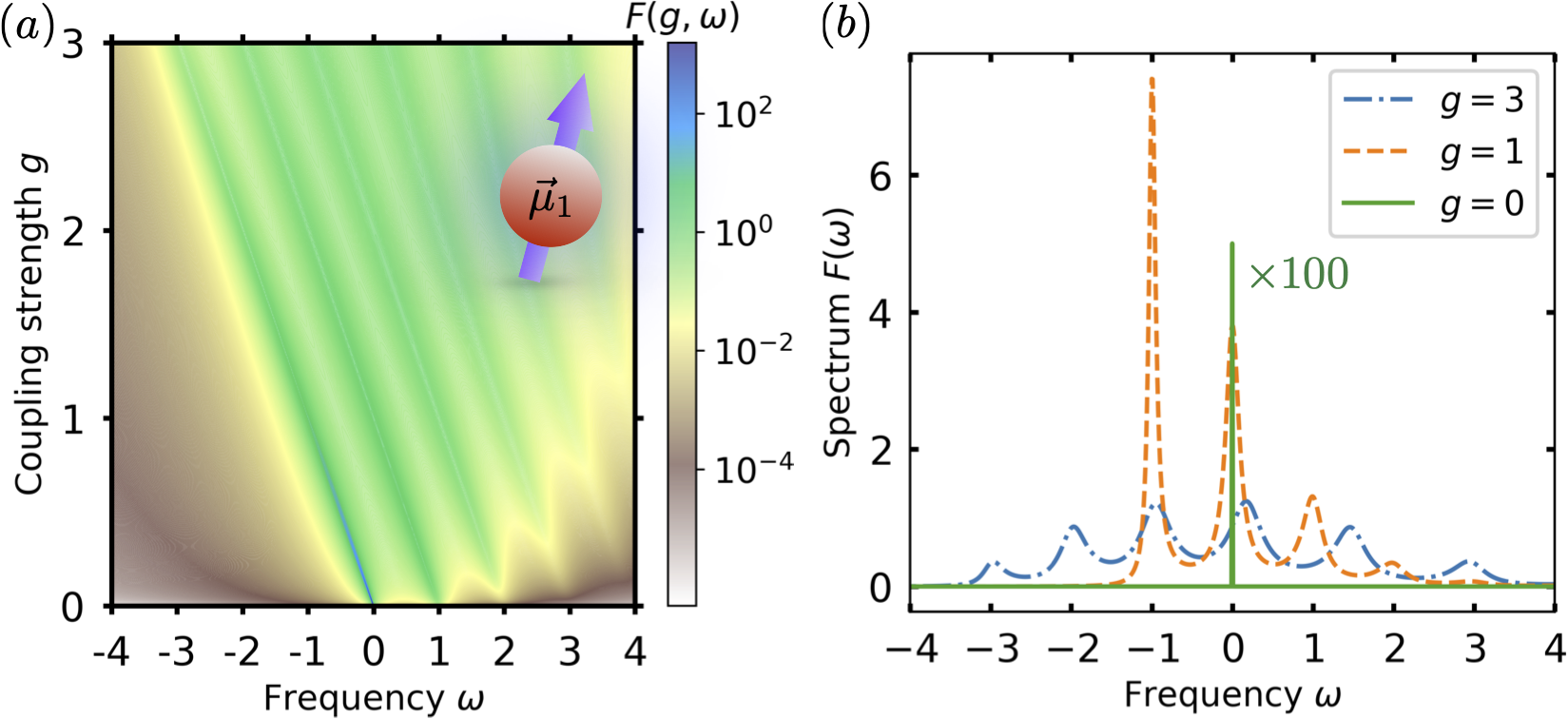} 
\caption{(color online) (a) Contour plot of the absorption spectrum $F(g, \omega)$ for a monomer as a function of bath coupling strength $g$ and frequency $\omega$. The color scale represents the intensity of the absorption spectrum, with the electronic transition frequency set at $\omega_0 = 0$. (b) One-dimensional spectra extracted from the contour plot for specific values of bath coupling strength: $g = 0$ (green solid line, scaled by 100 for visibility), $g = 1$ (orange dashed line), and $g = 3$ (blue dash-dotted line). The spectra are plotted in units of $\mu^2$. The hierarchy depth is fixed at 12, and other parameters are $\gamma = 0.05$ and $\Omega = 1$, unless otherwise specified. 
}
\label{fig:fig_monomer_coupling}
\end{figure}

Figure~\ref{fig:fig_monomer_coupling} illustrates the absorption spectrum $F(g, \omega)$ of a monomer [shown in the inset of Fig.~\ref{fig:fig_monomer_coupling}(a)] as a function of bath coupling strength $g$ and frequency $\omega$. 
As shown in Fig.~\ref{fig:fig_monomer_coupling}(a), when the monomer is decoupled from its bath (i.e., $g = 0$), a single sharp peak is observed at $\omega = \omega_1^{\rm (L)} = 0$ [see Eq.~(\ref{eq:1D_eigenvalue}) or  Table~\ref{table:obc} with $j=N=1$], representing the primary transition frequency of the monomer. As the bath coupling strength $g$ increases, the peak frequency shifts continuously towards negative values, and the spectral amplitude decreases. Additionally, new peaks appear, which are induced by  the vibrational bath and correspond to higher frequencies. These peaks converge towards $\omega \approx 1, 2, 3, 4, \dots$ for small $g$, indicating the emergence of vibrational bath modes within the system. 
The presence of multiple peaks indicates strong aggregate-bath coupling, resulting in memory effects, where the bath retains information and feeds it back into the quantum aggregate, thereby giving rise to non-Markovian behavior. 
Figure~\ref{fig:fig_monomer_coupling}(b) presents the one-dimensional spectra extracted from the contour plot in Fig.~\ref{fig:fig_monomer_coupling}(a) for three specific values of the bath coupling strength: $g = 0,\,1,\,3$. At $g = 0$, the spectrum exhibits a single sharp peak at $\omega = 0$, which is significantly more intense than the spectra for higher bath coupling strengths (scaled by a factor of 100). As $g$ increases to $g = 1$, the peak shifts to $\omega \approx -1$, and additional peaks emerge, reflecting the influence of the coupling on the absorption characteristics. At $g = 3$, the spectrum becomes more complex, with multiple peaks distributed across a wider frequency range, indicating that strong coupling significantly alters the absorption properties of the monomer, resulting in more pronounced vibrational resonances.

\begin{figure}
\centering
  \includegraphics[width=1\columnwidth]{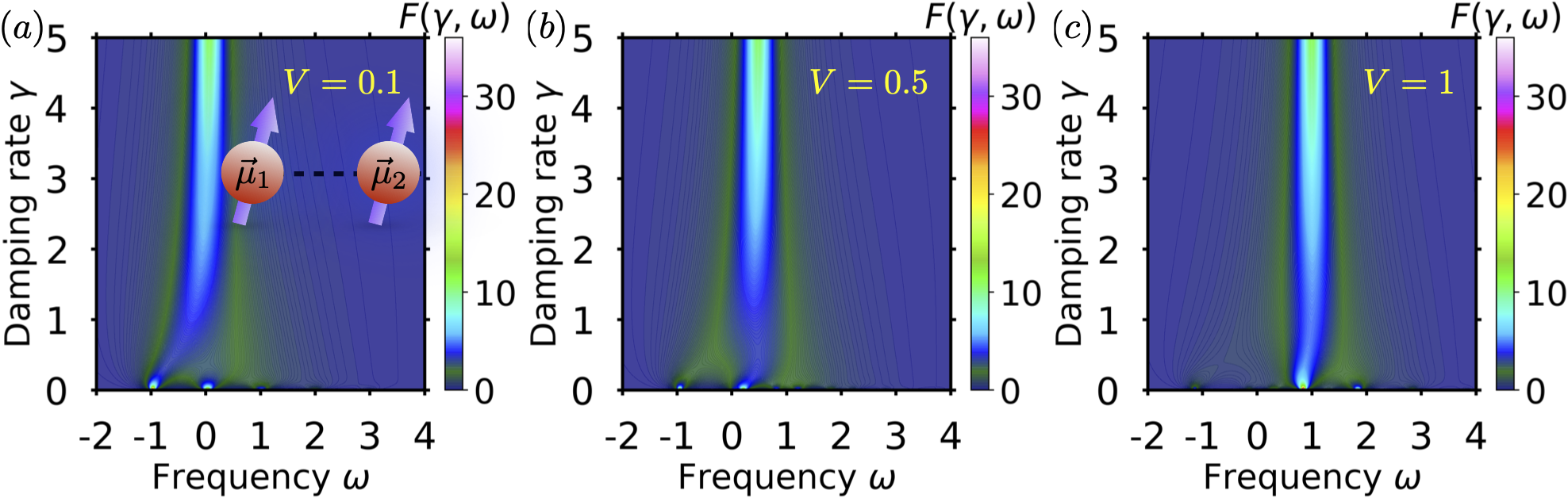} 
\caption{(color online) Absorption spectrum $F(\gamma,\omega)$ of a dimer as a function of the dissipation rate $\gamma$, shown for dipole-dipole interaction strengths of (a) $V = 0.1$, (b) $V = 0.5$, and (c) $V = 1$ between two monomers. The spectral peaks at large $\gamma$, consistent with the values predicted by Eq.~(\ref{eq:1D_eigenvalue}), are located at $\omega \approx \omega_1^{(L)} = 0,\, 0.1,\, 0.5$ for panels (a), (b), and (c), respectively. The hierarchy depth is set to 12, with the angle between the dipoles $\theta = 0$, and a bath coupling strength of $g = 1$. We set $\Omega=1$ as the energy unit. All other parameters are consistent with those used in Fig.~\ref{fig:fig_monomer}.
}
\label{fig:fig_dimerDamping}
\end{figure}

\subsubsection{Spectral analysis of a dimer}

We next study the dimer depicted in the inset of Fig.~\ref{fig:fig_dimerDamping}(a) and the dipoles are oriented in parallel ($\theta=0$). The calculated spectra are shown in Fig.~\ref{fig:fig_dimerDamping}. 
For a weak dipole-dipole interaction of $V = 0.1$ in Fig.~\ref{fig:fig_dimerDamping}(a), a peak at $\omega\approx\omega_{1,V=0.1}^{\rm (L)} = 0.1$, which is consistent with the value predicted by Eq.~(\ref{eq:1D_eigenvalue}) (see Table~\ref{table:obc} as well, where $V=1$ is used instead), is observed for a high dissipation rate $\gamma$ (e.g., $\gg 1$). 
While Eq.~(\ref{eq:1D_eigenvalue}) assumes $g = 0$, the result here uses $g = 1$. Despite this difference, Fig.~\ref{fig:fig_dimerDamping}(a) shows no noticeable deviation. 
This peak starts to split at $\gamma\approx 1$, indicating non-Markvoian features at small $\gamma$. The peak positions for $\gamma=0$ are almost the same as those of a monomer, while the associated amplitudes are enhanced. This enhancement arises because, for two uncoupled monomers with $V=0$, the amplitude is twice that of a single monomer. When the dipole-dipole interaction is increased to $V=0.5$ in Fig.~\ref{fig:fig_dimerDamping}(b) and $V=1.0$ in Fig.~\ref{fig:fig_dimerDamping}(c), the peak position at large $\gamma$ shifts to $\omega\approx\omega_{1,V=0.5}^{\rm (L)}=0.5$ and $\omega\approx\omega_{1,V=1}^{\rm (L)}=1$ respectively, in accordance with Eq.~(\ref{eq:1D_eigenvalue}) or Table~\ref{table:obc} with $j=1$ and $N=2$. Notably, in Fig.~\ref{fig:fig_dimerDamping}(c), the maximum spectral amplitude occurs at $\omega\lesssim 1.0$ when $\gamma=0$, in contrast to the monotonically decreasing amplitude with $\omega$ observed for a monomer in Fig.~\ref{fig:fig_monomer}(a) and Fig.~\ref{fig:fig_monomer}(b) or a weakly-coupled dimer as exemplified in Fig.~\ref{fig:fig_dimerDamping}(a). 
Comparing Figs.~\ref{fig:fig_dimerDamping}(a)-\ref{fig:fig_dimerDamping}(c), we further notice that the increase of the dipole-dipole interaction $V$ raises the transition frequency and shortens the system characteristic time, compared to a given bath correlation time $1/\gamma$ (e.g., for $\gamma \leq 1$). This suggests a potentially greater flow of  information from the bath back to the system, thereby enhancing non-Markovian effects. 
Alternatively, this can be understood as follows: an increase in the dipole-dipole interaction within a dimer aggregate, for a fixed dissipation, reduces the ratio $\gamma/V$. As a result, the relative impact of dissipation decreases, allowing more information to flow from the bath back to the system, leading to memory effects and, consequently, non-Markovian behavior.

\begin{figure}
\centering
  \includegraphics[width=1\columnwidth]{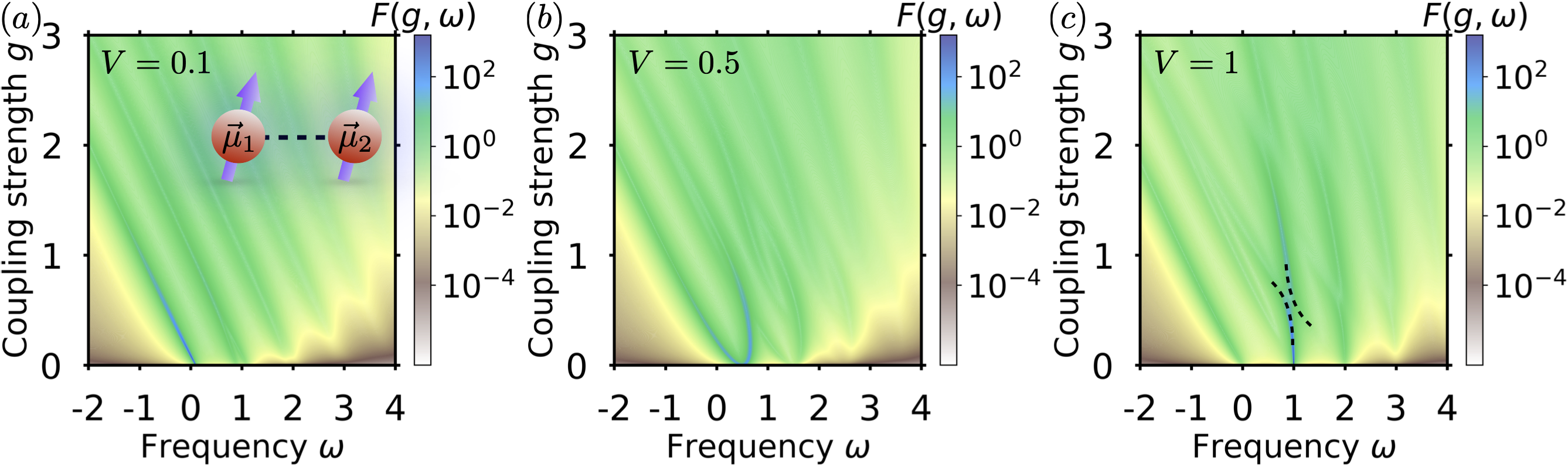} 
\caption{(color online) Absorption spectrum $F(g, \omega)$ of a dimer as a function of bath coupling strength $g$, shown for dipole-dipole interaction strengths of (a) $V = 0.1$, (b) $V = 0.5$, and (c) $V = 1$ between two monomers. 
The anti-crossing in (c) appears near $(\omega, g) \simeq (0.9, 0.7)$, with dashed curves offering clear visual guidance.
The logarithmic color scale highlights the absorption intensity across the frequency range $\omega$, with the energy unit normalized to $\Omega = 1$. The transition frequency is fixed at $\omega_0 = 0$. Other parameters, including the damping rate $\gamma = 0.05$ and hierarchy depth of 12, are consistent with those used in Fig.~\ref{fig:fig_monomer_coupling}. 
}
\label{fig:fig_dimerCoupling}
\end{figure}

Figure~\ref{fig:fig_dimerCoupling} illustrates the absorption spectrum $F(g, \omega)$ of a dimer [illustrated in the inset of Fig.~\ref{fig:fig_dimerCoupling}(a)] as a function of bath coupling strength $g$ for three distinct dipole-dipole interaction strengths: $V = 0.1$ [Fig.~\ref{fig:fig_dimerCoupling}(a)], $V = 0.5$ [Fig.~\ref{fig:fig_dimerCoupling}(b)], and $V = 1.0$ [Fig.~\ref{fig:fig_dimerCoupling}(c)].
At the lowest interaction strength of $V = 0.1$ in Fig.~\ref{fig:fig_dimerCoupling}(a), the absorption spectrum is characterized by broad and smooth features, with a peak emerging at frequency $\omega = \omega_{1,V=0.1}^{\rm (L)} = 0.1$ when $g = 0$ [see Eq.~(\ref{eq:1D_eigenvalue}) or Table~\ref{table:obc} where $V=1$ is considered instead]. As the bath coupling strength $g$ increases, the spectrum shifts gradually, and new peaks emerge, but the features remain broad, indicating weak dipole-dipole interactions. 
Increasing the interaction strength to $V = 0.5$ [Fig.~\ref{fig:fig_dimerCoupling}(b)] results in more distinct and structured absorption features. The initial peak at $\omega = \omega_{1,V=0.5}^{\rm (L)} = 0.5$, observed at $g = 0$, shifts and eventually splits as $g$ increases. 
At the highest interaction strength of $V = 1.0$ [Fig.~\ref{fig:fig_dimerCoupling}(c)], the absorption spectrum displays highly pronounced and well-defined peaks. 
In addition to the peak at $\omega=\omega_{1,V=1}^{(L)}=1$ when $g=0$, notably, anti-crossing behavior is observed at points such as $(\omega, g) \simeq (0.9, 0.7)$ indicated by the dashed curves for visual reference. 
This anti-crossing arises from energy splitting induced by strong dipole-dipole interactions, which reduces the relative effect of dissipation (represented by $\gamma/V$) and enhances non-Markovian behavior.

\begin{figure}
\centering
  \includegraphics[width=1\columnwidth]{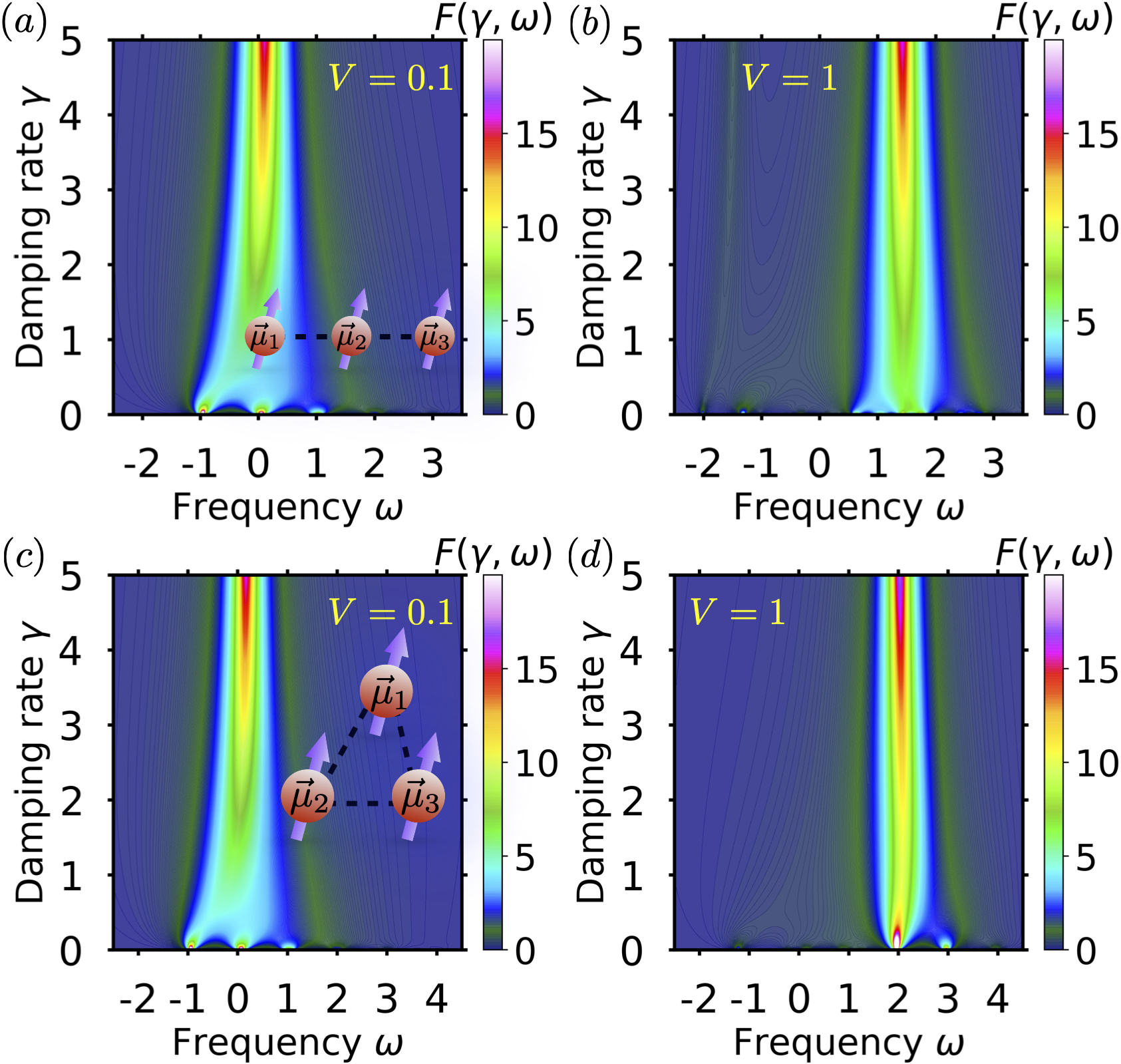} 
\caption{(color online) Absorption spectrum $F(\gamma,\omega)$ for a trimer in both linear (a, b) and ring (c, d) geometries, plotted as a function of dissipation rate $\gamma$. The dipole-dipole interaction between adjacent monomers is $V = 0.1$ in (a, c) and $V = 1$ in (b, d). In the linear configuration, the primary spectral peaks at large $\gamma$ are located at $\omega \approx \omega_1^{(L)} = \sqrt{2}/10 \approx 0.1414$ (panel a) and $\sqrt{2} \approx 1.414$ (panel b). An additional, weaker spectral peak is observed at $\omega \approx \omega_3^{(L)} = -\sqrt{2}\approx -1.414$ (panel b). 
For the ring configuration, the corresponding peaks are $\omega \approx \omega_3^{(C)} = 0.2,\,2$ in panels (c) and (d), respectively. The dipoles are aligned at an angle of $\theta = 0$. We set $\Omega=1$ as the energy unit. All other parameters are consistent with those used in Fig.~\ref{fig:fig_monomer}. 
}
\label{fig:fig_trimerDamping}
\end{figure}

\subsection{Configuration-Associated Spectral Signatures}
\label{IVC}

\subsubsection{Spectral response to vibrational dissipation}

The spectra for a linear trimer, shown in the inset of Fig.~\ref{fig:fig_trimerDamping}(a), are presented in Figs.~\ref{fig:fig_trimerDamping}(a) and \ref{fig:fig_trimerDamping}(b). In contrast, the spectra for a ring trimer, illustrated in the inset of Fig.~\ref{fig:fig_trimerDamping}(c) under periodic boundary conditions (see Sec.~\ref{sec:anaylitics}), are shown in Figs.~\ref{fig:fig_trimerDamping}(c) and \ref{fig:fig_trimerDamping}(d). Each set of figures corresponds to two different values of the dipole-dipole interaction, providing a comparison between the two configurations. 
In the case of the linear trimer with a weak dipole-dipole interaction of $V = 0.1$ shown in Fig.~\ref{fig:fig_trimerDamping}(a), a single peak is observed at $\omega \approx \omega_{1,V=0.1}^{\rm (L)} = \sqrt{2}/10 \approx 0.1414$ for large dissipation $\gamma$. Note that the secondary peak at $\omega \approx \omega_{3,V=0.1}^{(L)} = -\sqrt{2}/10\approx -0.1414$, which has a relatively weaker amplitude (see Table~\ref{table:obc} for $j=N=3$, where $V=1$ instead), falls within the bandwidth of the primary peak, rendering it indistinguishable. 
Reducing $\gamma$ leads to the splitting of this peak into multiple peaks at frequencies nearly identical to those of a monomer at $\gamma=0$ [see Figs.~\ref{fig:fig_monomer}(a) and \ref{fig:fig_monomer}(b)], indicating a transition from Markovian to non-Markovian regimes. 
When increasing $V$, besides the shift of the peak to $\omega \approx \omega_{1,V=1}^{\rm (L)}=\sqrt{2}\approx 1.414$ in Fig.~\ref{fig:fig_trimerDamping}(b), there is an additional peak with a relatively weak amplitude at $\omega\approx\omega_{3,V=1}^{\rm (L)}=-\sqrt{2}\approx -1.414$ in Fig.~\ref{fig:fig_trimerDamping}(b) for strong dissipation (e.g., $\gamma\sim 5$). %
Regarding the ring trimer, the spectra in Figs.~\ref{fig:fig_trimerDamping}(c) and \ref{fig:fig_trimerDamping}(d) exhibit peaks at $\omega \approx \omega_{3,V=0.1}^{\rm (C)}=0.2$ and $\omega \approx \omega_{3,V=1}^{\rm (C)}=2$, respectively, at large $\gamma$ [see Eq.~(\ref{eq:2D_eigenvalue}) or Table~\ref{table:pbc}]. These peaks exhibit behaviors similar to those in the dimer case shown in Fig.~\ref{fig:fig_dimerDamping}, but with distinct spectral amplitude intensities, as $\gamma$ decreases or $V$ increases. Importantly, the comparison of these spectra in Fig.~\ref{fig:fig_trimerDamping} demonstrates that one can distinguish a ring trimer from a linear trimer.

\begin{figure}
\centering
  \includegraphics[width=1\columnwidth]{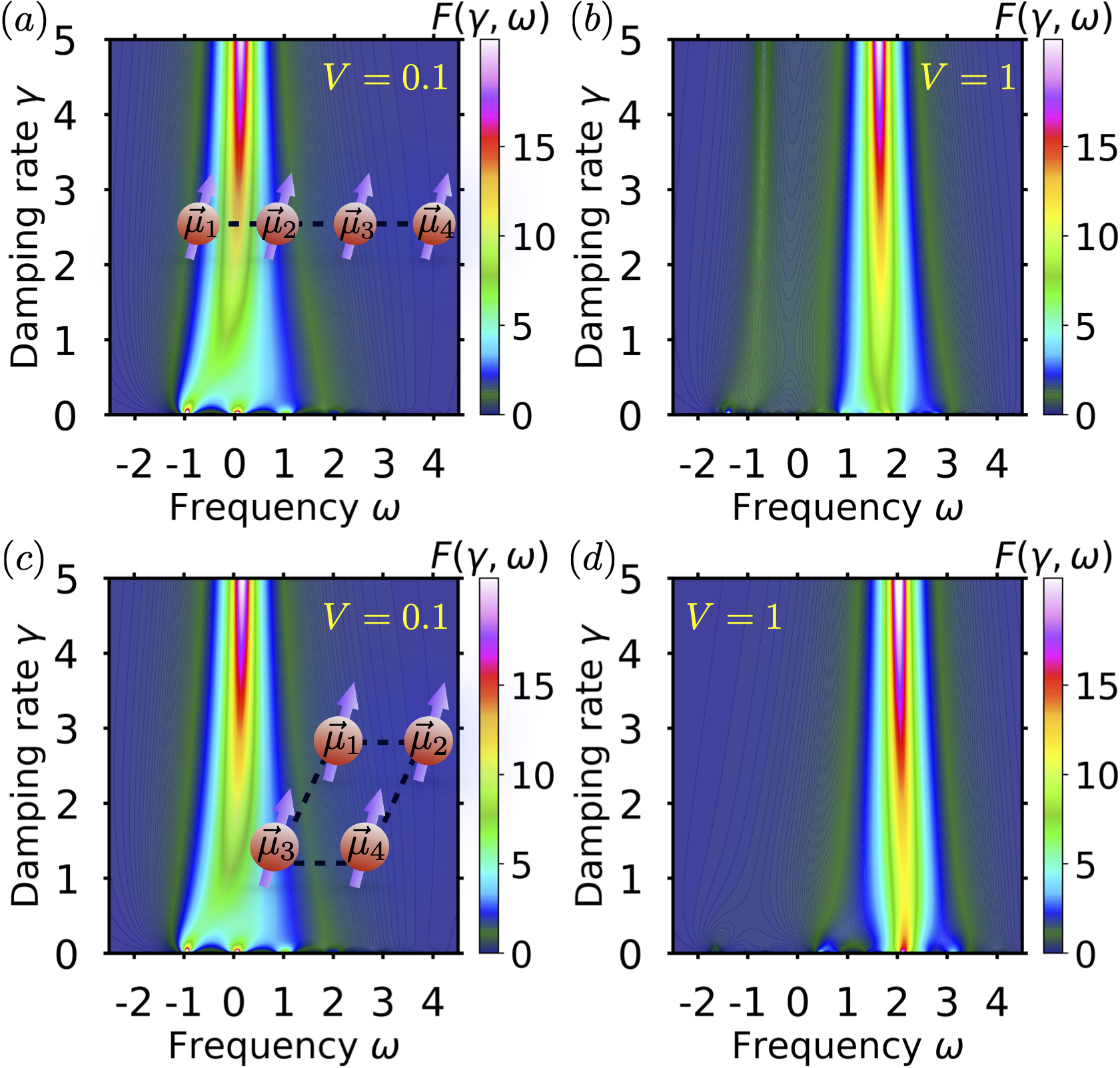} 
\caption{(color online) Absorption spectrum $F(\gamma,\omega)$ for a tetramer in both linear (a, b) and ring (c, d) geometries, plotted as a function of dissipation rate $\gamma$. The dipole-dipole interaction between adjacent monomers is $V = 0.1$ in (a, c) and $V = 1$ in (c, d). In the linear configuration, the primary spectral peaks at large $\gamma$ are located at $\omega \approx \omega_1^{(L)} = (\sqrt{5}+1)/20\approx 0.1618$ (panel a) and $(\sqrt{5}+1)/2\approx1.618$ (panel b). An additional, weaker spectral peak is observed at $\omega \approx \omega_3^{(L)} = -(\sqrt{5}-1)/2 \approx -0.618$ (panel b). 
For the ring configuration, the corresponding peaks are $\omega \approx \omega_4^{(C)} = 0.2,\,2$ in panels (c) and (d), respectively. The dipoles are aligned at an angle of $\theta = 0$. We set $\Omega=1$ as the energy unit. All other parameters are consistent with those used in Fig.~\ref{fig:fig_monomer}.
}
\label{fig:fig_tetramerDamping}
\end{figure}

A linear tetramer, shown in the inset of Fig.~\ref{fig:fig_tetramerDamping}(a), has a spectrum displayed in Fig.~\ref{fig:fig_tetramerDamping}(a) for $V=0.1$, which features a peak at $\omega \approx \omega_{1,V=0.1}^{\rm (L)} = (\sqrt{5}+1)/20 \approx 0.1618$ for strong dissipation (e.g., $\gamma\sim 5$) [see Eq.~(\ref{eq:1D_eigenvalue}) and Table~\ref{table:obc} for $j=1$ and $N=4$]. 
Further peak splittings are induced not only by the decrease in $\gamma$, but also by the increase in $V$, resembling the behavior seen in the linear trimer case, but differing in spectral intensity. The former reflects bath-induced vibrational transitions, while the latter represents the splitting of the system's electronic states.   
For example, for large $\gamma$, the increase of $V$ splits the peak into two peaks: $\omega\approx\omega_{1,V=1}^{\rm (L)}=(\sqrt{5}+1)/2 \approx1.618$ with strong amplitude and $\omega\approx \omega_{3,V=1}^{\rm (L)}=-(\sqrt{5}-1)/2 \approx -0.618$ with weak amplitude in Fig.~\ref{fig:fig_tetramerDamping}(b). The distance between these two peaks increases from $2.236$ when decreasing $\gamma$ from infinity, which is smaller than $2.828$ for a trimer shown in Fig.~\ref{fig:fig_trimerDamping}(b). 
These unique spectral features can help distinguish between linear trimers and tetramers.
In contrast to the linear configuration, the spectra of a ring tetramer, depicted in the inset of Fig.~\ref{fig:fig_tetramerDamping}(c) and presented in Figs.~\ref{fig:fig_tetramerDamping}(c) and \ref{fig:fig_tetramerDamping}(d), exhibit behaviors similar to but distinct from those of a ring trimer, shown in the inset of Fig.~\ref{fig:fig_trimerDamping}(c) and illustrated in Figs.~\ref{fig:fig_trimerDamping}(c) and \ref{fig:fig_trimerDamping}(d). 
Therefore, an absorption spectrum with an appropriate dipole-dipole interaction could be used to distinguish a linear tetramer from a ring one.

\subsubsection{Spectral response to aggregate-bath coupling strength}

\begin{figure}
\centering
  \includegraphics[width=1\columnwidth]{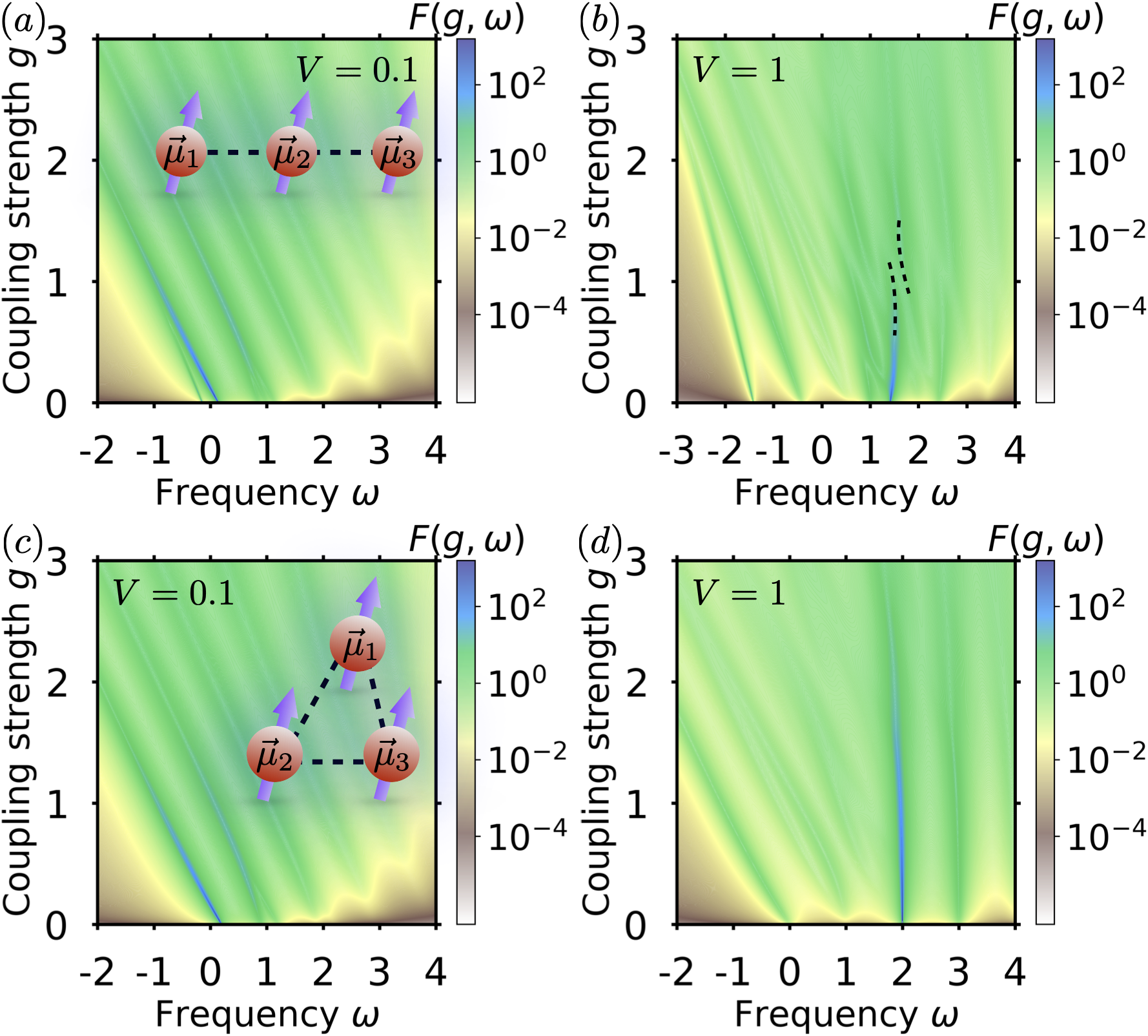} 
\caption{(color online) Absorption spectrum $F(g, \omega)$ of a trimer system in two different geometries: linear [top row: (a), (b)] and ring [bottom row: (c), (d)], as a function of bath coupling strength $g$. 
The anti-crossing in (b) appears near $(\omega, g) \simeq (1.5, 1)$, with dashed curves offering clear visual guidance.
The dipole-dipole interaction strength $V$ between adjacent monomers is set to (a, c) $V = 0.1$ and (b, d) $V = 1$. The energy unit is normalized to $\Omega = 1$. All other parameters, including the damping rate $\gamma$ and hierarchy depth, align with those used in Fig.~\ref{fig:fig_monomer_coupling}. 
}
\label{fig:fig_trimerCoupling}
\end{figure}

Figure~\ref{fig:fig_trimerCoupling} displays the absorption spectrum $F(g, \omega)$ of a trimer system, illustrated in the inset of Fig.~\ref{fig:fig_trimerCoupling}(a), as a function of bath coupling strength $g$ and frequency $\omega$, for both linear [top row: panels (a), (b)] and ring [bottom row: panels (c), (d)] geometries, with varying dipole-dipole interaction strengths. 
In the linear trimer configuration with a weak dipole-dipole interaction $V = 0.1$ [Fig.~\ref{fig:fig_trimerCoupling}(a)], the spectrum initially exhibits two symmetric peaks at $\omega = \omega_{1,V=0.1}^{\rm (L)} =\sqrt{2}/10 \approx 0.1414$ with strong amplitude and $\omega = \omega_{3,V=0.1}^{\rm (L)} =-\sqrt{2}/10 \approx -0.1414$ with weak amplitude when $g = 0$ [see Eq.~(\ref{eq:1D_eigenvalue}) and Table~\ref{table:obc}]. 
As the bath coupling strength $g$ increases, these peaks begin to merge, with the strong peak shifting towards negative frequencies and the weak peak merging into the spectrum. 
%
With an increase in the dipole-dipole interaction to $V = 1.0$ [Fig.~\ref{fig:fig_trimerCoupling}(b)], the spectrum shows two symmetric peaks at $\omega \approx \omega_{1,V=1}^{\rm (L)} =\sqrt{2} \approx 1.414$ with strong amplitude and $\omega \approx \omega_{3,V=1}^{\rm (L)} =-\sqrt{2} \approx -1.414$ with weak amplitude (when $g=0$), along with an asymmetric peak at $\omega \approx -0.4$ (for small $g >0$). As $g$ increases, the asymmetric peak merge with $\omega_{3,V=1}^{\rm (L)}$. 
Notably, an anti-crossing is observed at $(\omega, g) \approx (1.5, 1)$, as indicated by the dashed curves for visual reference, highlighting the significant energy level interactions and providing further evidence of non-Markovian effects.  

In contrast, the spectra for the ring trimer configuration, shown in the inset of Fig.~\ref{fig:fig_trimerCoupling}(c), display distinct behavior. For a weak interaction $V = 0.1$ [Fig.~\ref{fig:fig_trimerCoupling}(c)], a single peak appears at $\omega = \omega_{3,V=0.1}^{\rm (C)} = 0.2$ when $g = 0$ [see Eq.~(\ref{eq:2D_eigenvalue}) or Table~\ref{table:pbc}]. 
At $V = 1$ [Fig.~\ref{fig:fig_trimerCoupling}(d)], the ring trimer spectrum exhibits a peak at $\omega = \omega_{3,V=1}^{\rm (C)} = 2$ when $g=0$. The increase in bath coupling strength $g$ does not lead to merging of peaks, maintaining a more stable and isolated spectral structure, which contrasts with the linear trimer and further highlights the role of geometry in influencing the non-Markovian effects.

\begin{figure}
\centering
  \includegraphics[width=1\columnwidth]{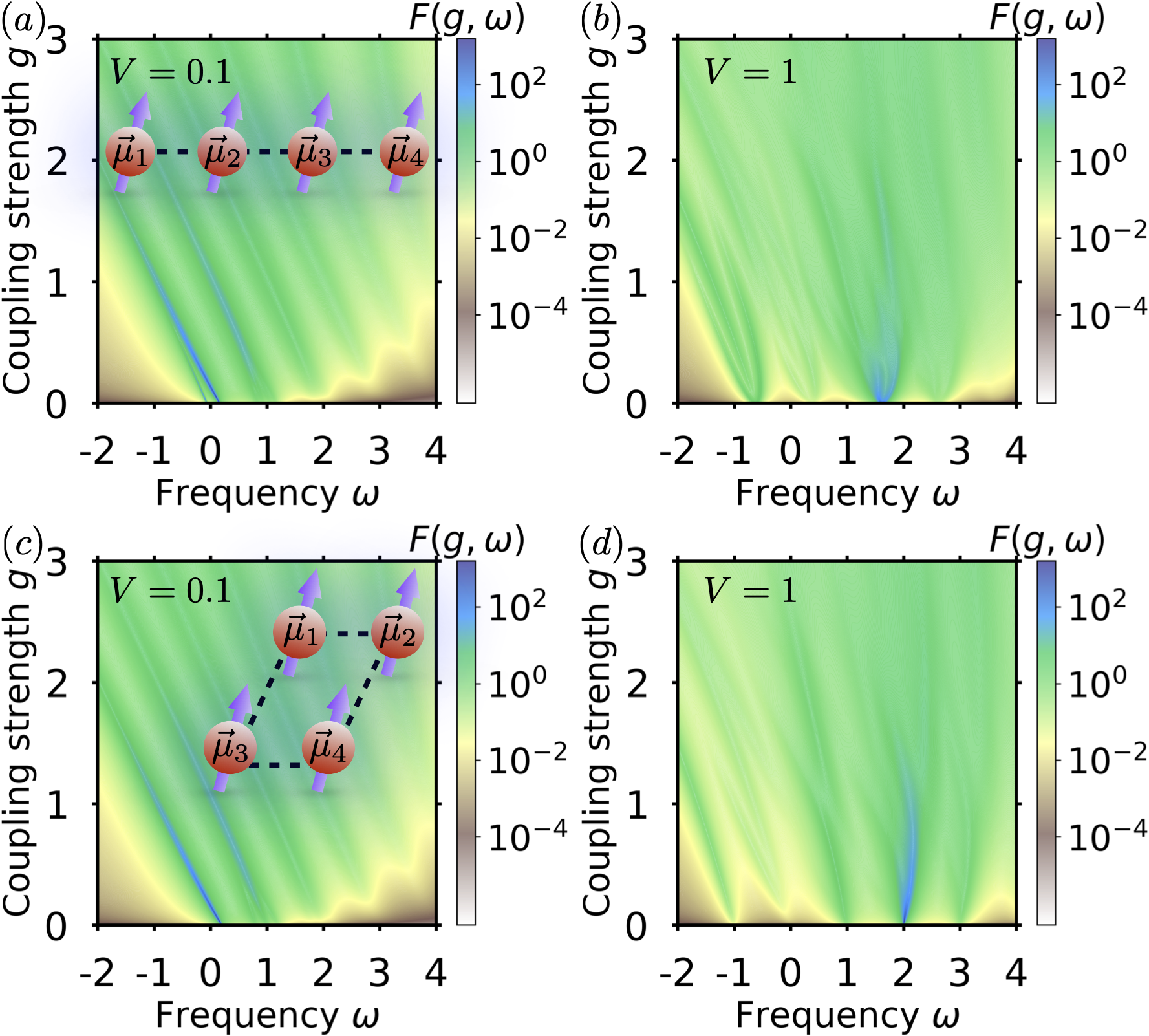} 
\caption{(color online) Absorption spectrum $F(g, \omega)$ of a tetramer system in two different geometries: linear [top row: (a), (b)] and ring [bottom row: (c), (d)], as a function of the bath coupling strength $g$ for various dipole-dipole interaction strengths between adjacent monomers: (a, c) $V = 0.1$ and (b, d) $V = 1$.  
The energy unit is set to $\Omega = 1$. All other parameters, including the damping rate $\gamma$ and hierarchy depth, are consistent with those used in Fig.~\ref{fig:fig_monomer_coupling}. 
}
\label{fig:fig_tetramerCoupling}
\end{figure}

Figure~\ref{fig:fig_tetramerCoupling} illustrates the absorption spectrum $F(g, \omega)$ of a tetramer system, shown in the inset of Fig.~\ref{fig:fig_tetramerCoupling}(a), for varying dipole-dipole interaction strengths. The spectra are compared for linear [top row: panels (a), (b)] and ring [bottom row: panels (c), (d)] geometries, as a function of bath coupling strength $g$ and frequency $\omega$.   
In the linear tetramer at a weak dipole-dipole interaction $V = 0.1$ [Fig.~\ref{fig:fig_tetramerCoupling}(a)], the spectrum initially exhibits two separate peaks at $\omega = \omega_{1,V=0.1}^{\rm (L)} = (\sqrt{5}+1)/20 \approx 0.1618$ with strong amplitude and  $\omega = \omega_{3,V=0.1}^{\rm (L)} = -(\sqrt{5}-1)/20 \approx -0.0618$ with weak amplitude when $g = 0$. As the bath coupling strength $g$ increases, these peaks gradually merge. 
At the interaction strength $V = 1$ [Fig.~\ref{fig:fig_tetramerCoupling}(b)], the peaks at $\omega \approx \omega_{1,V=1}^{\rm (L)} = (\sqrt{5}+1)/2 \approx 1.618$ with strong amplitude and $\omega \approx \omega_{3,V=1}^{\rm (L)} = -(\sqrt{5}-1)/2 \approx -0.618$ with weak amplitude splits as $g$ increases. This splitting reflects 
significant interactions between the tetramer components and their environment.

In the ring tetramer with a weak dipole-dipole interaction of $V = 0.1$ [Fig.~\ref{fig:fig_tetramerCoupling}(c)], the spectrum features a peak at $\omega = \omega_{4,V=0.1}^{(C)} = 0.2$ for $g = 0$ and exhibits broad and smooth characteristics.  
As $g$ increases, the peaks shift slightly, but no significant splitting  
occurs. The stability of the spectrum in the ring configuration suggests that non-Markovian effects are less pronounced, likely due to the symmetry and closed-loop structure, which stabilizes the system's response to environmental interactions.
As the dipole-dipole interaction strength increases to $V = 1$ [Fig.~\ref{fig:fig_tetramerCoupling}(d)], the spectrum shows a peak at $\omega = \omega_{4,V=1}^{(C)} = 2$ for $g = 0$.  
The lack of splitting or emerging at higher dipole-dipole interaction strengths suggests that the symmetry of the ring contributes to a more stable spectral structure compared to the linear tetramer.

Finally, we summarize the impact of the dipole-dipole interaction $V$ on the absorption spectrum as demonstrated by our results (see e.g., Figs.~\ref{fig:fig_dimerDamping}, \ref{fig:fig_trimerDamping}, and \ref{fig:fig_tetramerDamping}). At high dissipation rates ($\gamma$), $V$ induces peak splitting in the linear trimer [Fig.~\ref{fig:fig_trimerDamping}(b)] and tetramer [Fig.~\ref{fig:fig_tetramerDamping}(b)], with one peak exhibiting strong amplitude and another showing weak amplitude. This splitting, which may indicate enhanced non-Markovian effects, is absent in the dimer (Fig.~\ref{fig:fig_dimerDamping}), ring trimer [Figs.~\ref{fig:fig_trimerDamping}(c) and \ref{fig:fig_trimerDamping}(d)], and ring tetramer  [Figs.~\ref{fig:fig_tetramerDamping}(c) and \ref{fig:fig_tetramerDamping}(d)], where the spectral features remain more stable. When $\gamma$ is low and $V$ is small (e.g., $V = 0.1$ in Figs.~\ref{fig:fig_dimerDamping}, \ref{fig:fig_trimerDamping}, and \ref{fig:fig_tetramerDamping}), the spectral amplitude decreases progressively from the first peak onward, with the strongest amplitude at the initial peak, resembling the behavior observed in monomers and dimers.  
In contrast, for larger values of $V$, the redistribution of spectral amplitude leads to a shift in maximum intensity away from the first peak, highlighting the influence of more significant non-Markovian effects.
Note that, in addition to the peak splitting with strong and weak amplitudes induced by the strong dipole-dipole interaction in Figs.~\ref{fig:fig_trimerDamping}(b) and \ref{fig:fig_tetramerDamping}(b) for the spectrum $F(\gamma, \omega)$, similar peak splitting with strong and weak amplitudes is also observed in Figs.~\ref{fig:fig_trimerCoupling} and \ref{fig:fig_tetramerCoupling} for $F(g, \omega)$, as expected in the limit $\gamma \rightarrow \infty$ or $g \rightarrow 0$.

\section{Discussions and conclusions \label{sec:conclusion}}

In this work, we introduced absorption spectra as an innovative and experimentally accessible tool to probe the Markovian to Non-Markovian transition. By systematically analyzing quantum aggregates with varying geometries, we identified distinct spectral signatures—such as peak splitting, merging, and shifting—that directly indicate non-Markovian behavior. These features are intricately tied to key physical parameters, including dissipation rate, aggregate-bath coupling strength, and intra-aggregate dipole-dipole interactions. 

We performed numerical calculations of absorption spectra using Laplace-domain hierarchical equations for linear and ring aggregates coupled to a vibrational bath with a Lorentzian noise spectrum at zero temperature, noting that extensions to finite temperatures are straightforward based on prior studies~\cite{Yu04pra,RitschelEisfeld15jcp}. Our results reveal that reducing the dissipation rate induces spectral peak splitting, marking the transition from Markovian to non-Markovian regimes, while enhanced dipole-dipole interactions also drive peak splitting through a distinct mechanism. Increasing the aggregate-bath coupling strength leads to merging of initially symmetric or asymmetric peaks under weak dipole-dipole interactions, but causes peak splitting under strong interactions. Furthermore, absorption spectra clearly differentiate linear and ring geometries: linear aggregates, with reduced symmetry, exhibit intricate peak splitting and merging that reflect stronger non-Markovian effects under varying conditions, whereas ring geometries, characterized by inherent symmetry and periodic boundary conditions, maintain stable and isolated peaks, demonstrating lower susceptibility to non-Markovian influences. These findings establish a comprehensive framework linking spectral features to non-Markovian behavior and underscore the potential of absorption spectra as a practical tool for exploring and controlling memory effects in diverse quantum systems.

In our analysis, we set the central frequency $\Omega = 1$ as a reference point to establish a consistent energy scale, enabling a systematic examination of how variations in spectral width $\gamma$ affect the transition between Markovian and non-Markovian behavior. The Markovianity of the dynamics depends primarily on the ratio of $\gamma$ to $\Omega$, rather than the absolute value of \( \Omega \). When $\gamma$ is large relative to $\Omega$, the spectral density broadens, leading to rapid decay of the bath correlation function and Markovian dynamics. Conversely, a small $\gamma$ narrows the spectral density around $\Omega$, increasing bath coherence time, enhancing memory effects, and promoting non-Markovian behavior. Fixing $\Omega = 1$ provides a stable reference for observing the influence of spectral broadening on the behavior without a direct impact from $\Omega$ itself.

The transition from Markovian to non-Markovian behavior can be characterized by measurable thresholds, such as the aggregate-bath coupling strength or the spacing between vibration-induced peaks relative to their widths. In the Markovian regime, pure electronic peaks dominate when the coupling is zero or dissipation is extremely strong. However, as coupling increases or dissipation weakens, the peaks become influenced by the vibrational bath. If the spacing between these peaks remains smaller than their widths, the system retains Markovian characteristics. Once the spacing surpasses the widths, the peaks become distinct, signaling a shift to the non-Markovian regime. This transition is marked by reduced dissipation, stronger coupling, the emergence of memory effects, and more intricate interactions within the system.

Although our system operates with a Hermitian Hamiltonian and lacks exceptional points (EPs), there are conceptual parallels between the transitions we observe and those in non-Hermitian ${\cal PT}$-symmetric systems~\cite{GaikwadMurch24prl}. In open quantum systems governed by Lindblad dynamics, particularly in the absence of quantum jumps, effective non-Hermitian Hamiltonians can describe the dynamics~\cite{LiWhaley24prr,PeterYelin}. The ${\cal PT}$-symmetry phase transition, characterized by a Hamiltonian EP, resembles the Markovian to non-Markovian transition seen in our study. Exploring these connections further, including the impact of quantum jumps, will be a focus of future research.

Our study introduces a new approach to probing the Markovian to Non-Markovian transition in open quantum systems through absorption spectrum analysis, advancing beyond prior works. For instance, Ref.~\onlinecite{RitschelEisfeld15jcp} employed non-Markovian quantum state diffusion to compute temperature-dependent linear spectra of chromophore aggregates but did not explore the Markovian to Non-Markovian transition or the role of structured noise in shaping spectral features. Similarly, Ref.~\onlinecite{RodenStrunzEisfeld11JCP} focused on validating approximations in non-Markovian quantum state diffusion using pseudomode calculations, without leveraging spectral analysis to identify this transition. In contrast, our work systematically investigates distinct spectral features—such as peak splitting, shifting, and merging—that serve as direct and experimentally accessible indicators of the Markovian to Non-Markovian transition. By linking these features to aggregate-bath coupling, dissipation, and geometry, we provide configuration-specific insights into non-Markovian effects across various aggregate geometries, including monomers, dimers, trimers, and tetramers. 
A key innovation of our approach is the use of hierarchical algebraic equations in the frequency domain, which derive a hierarchy of linear equations directly in the Laplace domain. This method achieves finer-grained resolution of spectral features and memory effects compared to conventional time-domain techniques, offering deeper insights into the dynamics of complex quantum systems. 
Building on these advancements, our findings establish a versatile framework for studying non-Markovian behavior, paving the way for further exploration in quantum information, light-harvesting systems, and materials science.

Various measures have been developed to quantify non-Markovian effects in dynamical systems~\cite{BreuerLainePiilo09PRL,RivasHuelgaPlenio10PRL,LuWangSun10PRA,Luo10PRA,PollockModi18prl,LiWiseman18PhysRep,ShrikantMandayam23}. Among these, the trace distance measure~\cite{BreuerLainePiilo09PRL} is widely used, particularly in experimental studies of Markovian to non-Markovian transitions~\cite{LiuLiBreuerPiilo11nphys}. This measure identifies memory effects by tracking changes in the distinguishability between quantum states, where an increase in trace distance signals information backflow, which is a hallmark of non-Markovianity.
Although trace distance provides a general operational perspective, it abstracts memory effects without directly addressing the underlying physical mechanisms or environmental structures. In contrast, our spectral approach connects non-Markovianity to specific bath properties through observable spectral features, such as resolvable peaks in absorption spectra. These features arise under conditions of reduced dissipation or strong coupling to the vibrational
bath, offering an experimentally accessible way to probe the physical origins of non-Markovian behavior.
There is a natural connection between these two measures. The environmental memory effects that increase trace distance often correspond to the structural characteristics revealed in the spectrum. By focusing on the frequency domain, our spectral method highlights how dissipation strength, aggregate-bath coupling, intra-aggregate dipole-dipole interactions, 
and geometry influence non-Markovian transitions, providing a clear understanding of these dynamics. Unlike time-domain techniques, which often struggle to resolve complex features like multiple non-exponential oscillations, the spectral approach offers a direct and detailed representation of memory effects.
These two methods together provide complementary insights. While the trace distance measure offers a broad operational view of memory effects, the spectral approach delves into the
environmental structures and dynamics driving these effects. By combining these perspectives, our framework enhances the understanding of non-Markovian effects and provides a versatile and experimentally grounded method for exploring and controlling these phenomena. 

In open quantum systems, non-Markovianity can be characterized by either environmental spectral properties or the divisibility of the reduced system’s dynamics. A non-flat spectral density typically indicates non-Markovianity by introducing memory effects via structured environmental frequencies. Alternatively, non-Markovianity is sometimes defined by completely positive trace-preserving (CPTP) divisibility: if the system’s evolution can always be divided into CPTP maps, it is considered Markovian; otherwise, it is non-Markovian. Recent findings (e.g., Ref.~\onlinecite{MilzModi19PRL}) challenge the view that divisibility alone implies Markovianity, showing that even CPTP-divisible processes can display non-Markovian correlations. Our work examines non-Markovianity from a spectral perspective, focusing on how vibrational structures affect dynamics without necessarily implying a breakdown of CPTP-divisibility. By clarifying this distinction, we offer a nuanced interpretation of the non-Markovian features observed in our results across different frameworks.

Our proposal is not confined to molecular aggregates, such as Rydberg aggregates~\cite{Schonleber15prl} and excitonic systems in semiconductor quantum dots~\cite{Xu07science}, but extends to advanced quantum platforms. In molecular aggregates, external electric or magnetic fields can reshape energy landscapes, modulating aggregate-bath coupling, dissipation, and dipole-dipole interactions~\cite{FittipaldiSessoli19nmat}, while tailored environments provide control over dissipation processes. Beyond these systems, platforms like trapped ions and superconducting qubits offer promising opportunities to investigate spectral signatures of the Markovian to non-Markovian transition, expanding experimental possibilities and advancing the exploration of non-Markovian dynamics in diverse quantum systems.

Our findings on spectral signatures of the Markovian to non-Markovian transition can be directly tested using well-established experimental techniques. Linear absorption spectra, as a fundamental observable in molecular spectroscopy, can be readily obtained through standard broadband absorption measurements, where a tunable electromagnetic field probes the sample and the frequency-dependent absorption intensity is recorded~\cite{Fidder91JCP,Brixner04OL}. Additionally, ultrafast spectroscopic techniques such as pump-probe spectroscopy and two-dimensional electronic spectroscopy (2DES) provide deeper insights into transient spectral features, offering a direct means to observe bath-induced non-Markovian effects~\cite{Engel07Nature,Cho08,Jonas03}. These experimental techniques have been successfully employed to study excitonic coherence, bath-induced energy transfer, and non-Markovian relaxation in photosynthetic complexes and artificial aggregates~\cite{Lee07Science}. The theoretical predictions presented in this work, particularly regarding spectral peak splitting and its dependence on dissipation, aggregate-bath coupling, and dipole-dipole interactions, can thus be validated using these methods.

Our findings offer valuable insights into the geometry and electronic transitions of quantum aggregates while providing a practical framework for experimentally detecting and controlling the Markovian to non-Markovian transition in open quantum systems—an essential step toward advancing quantum technologies. Future research could extend our spectral analysis to phase-modulated nonlinear spectroscopy~\cite{Tekavec06JCP,Bruder15PRA,Osipov16PRA,LiEisfeld17pra} and examine the role of non-Markovian effects on quantum coherence and energy transfer in molecular systems, such as photosynthetic complexes, to deepen understanding of these fundamental processes. Additionally, exploring the interplay between non-Markovian dynamics and quantum error correction could improve error mitigation in quantum information processing and drive progress in quantum control and sensing technologies~\cite{WhiteHollenbergModi20ncommun,Hakoshima21pra,Ahn24aqt,RossiniDonvil23prl,ReichKoch15srep}.

\acknowledgements
Z.Z.L is grateful to Weijian Chen for valuable comments and suggestions.  
B.L. acknowledges support from the National Natural Science Foundation of China (Grant No. 12205066), the Guangdong Basic and Applied Basic Research Foundation (Grant No. 2024A1515011775), the Shenzhen Start-Up Research Funds (Grant No. BL20230925), and the start-up funding from the Harbin Institute of Technology, Shenzhen (Grant No. 20210134). C.T.Y. acknowledges support from the Guangdong Basic and Applied Basic Research Foundation (Grant No. 2214050004792), the National Natural Science Foundation of China (Grant No. 1217040938), and the Shenzhen Municipal Science and Technology projects (Grant No. 202001093000117).

\section*{Author Declarations}

\subsection*{Conflict of interest}
The authors have no conflicts to disclose.

\appendix

\section{Derivation of Eq.~(\ref{eq:hops_linear_laplace_nonoise}) \label{sec:AppendA}}

The calculation of the absorption spectrum is based on the solution to the hierarchy of equations in the Laplace domain~\cite{ZhangLiEisfeld17}. This hierarchy was formulated given that it could be less expensive to solve an algebraic equation than a differential equation when studying spectral properties. For the sake of completeness, in this appendix we provide details how Eq.~(\ref{eq:hops_linear_laplace_nonoise}) is derived. The coupled stochastic equations for an infinite hierarchy of pure states reads~\cite{SussEisfeldStrunz14prl,RitschelEisfeld15jcp} 
\begin{eqnarray}
\partial_t |\psi^{(\vec{k})}\rangle &=& (-iH_{\mathrm{sys}}-\sum_{n,j}k_{nj}w_{nj}+\sum_{n,j}L_n z_{nj}^{\ast}(t)) |\psi^{(\vec{k})}\rangle \notag\\
&&+\sum_{n,j}L_n k_{n,j}\alpha_{nj}(0) |\psi^{(\vec{k}-\vec{e}_{nj})}\rangle \notag\\
&&-\sum_{n,j}L_n^{\dagger} |\psi^{(\vec{k}+\vec{e}_{nj})}\rangle.
\label{eq:hops_linear}
\end{eqnarray}
Here a general form of bath correlation function for the coupling to the $n$th system operator $\alpha_n(t-s) = \sum_{j=1}^J \alpha_{nj}(t-s) = \sum_{j=1}^J k_{nj} e^{-w_{nj}(t-s)}$ is considered. 
Apply the Laplace transformation
\begin{eqnarray}
\mathcal{L}[f(t)] &=& \int_0^{\infty} dt e^{-s t} f(t),
\end{eqnarray}
to Eq.~(\ref{eq:hops_linear}), we have 
\begin{eqnarray}
|\psi^{(\vec{k})}(t=0)\rangle &=&
(s+iH_{\rm sys}+\sum_{n,j}k_{nj}\omega_{nj}) |\Psi^{(\vec{k})}(s)\rangle \notag\\
&&-\sum_{nj}L_n \mathcal{L}[z^{\ast}_{nj}(t) |\psi^{(\vec{k})}\rangle ] \notag \\
&&-\sum_{n,j}L_n k_{nj}\alpha_{nj}(0) |\Psi^{(\vec{k}-\vec{e}_{nj})}\rangle \notag\\
&&+\sum_{n,j}L_n^{\dagger} |\Psi^{(\vec{k}+\vec{e}_{nj})}\rangle, 
\label{eq:hops_linear_laplace}
\end{eqnarray}
where $z^*_{nj}(t)$ are complex Gaussian stochastic processes with the mean ${\rm E}[z_{nj}(t)]=0$ and correlations ${\rm E}[z_{nj}(t)z_{nj}(s)]=0$ and ${\rm E}[z_{nj}(t)z^*_{nj}(s)]=\alpha(t-s)$, and the Laplace transformation of a product of two functions is written formally as 
\begin{eqnarray}
\mathcal{L}[z^{\ast}_{nj}(t)|\psi^{(\vec{k})}\rangle] 
&=& \frac{1}{2\pi i} \int_{x-i\infty}^{x+i\infty} Z^{\ast}_{nj}(p) |\Psi^{(k)}(s-p)\rangle dp. \notag\\
\end{eqnarray}
The integration is done along the vertical line $Re[p]=x$ that lies entirely within the region of convergence of $Z^{\ast}_{nj}$. 
Due to the presence of this term, a compact form of the Laplace-domain equation becomes challengeable. However, when only considering the zeroth-order terms with respect to noise or studying the absorption spectrum where a stochastic noise plays no role~\cite{RitschelEisfeld15jcp}, the term $\mathcal{L}[z^{\ast}_{nj}(t)|\psi^{(\vec{k})}\rangle]$ doesn't contribute and therefore can be neglected. Therefore for the linear absorption and bath correlation function in Eq.~(\ref{eq:bcf}) (e.g., $J\rightarrow1$, $w_{nj}\rightarrow w_n$, and $k_{nj}\rightarrow g$) considered in this work and $|\psi^{(\vec{k})}(t=0)\rangle=|\Psi^{(\vec{k})}(s=0)\rangle$, Eq.~(\ref{eq:hops_linear_laplace}) becomes Eq.~(\ref{eq:hops_linear_laplace_nonoise}).

\end{document}